\documentclass[print]{jicspack}

\usepackage{enumerate}
\usepackage{graphicx}
\usepackage{subfigure}
\usepackage{amssymb}
\usepackage[lined,boxed,commentsnumbered, ruled]{algorithm2e}

\begin{document}

\setlength{\parindent}{10pt}

\begin{premaker}


\title{Extended Equal Service and Differentiated Service Models for Peer-to-Peer File Sharing}
\author[author1]{Jianwei Zhang},
\author[author2]{Yongchao Wang},
\author[author1]{Wei Xing\corauthref{cor1}},
\ead{wxing@zju.edu.cn}
\corauth[cor1]{Corresponding author.}
\author[author1]{Dongming Lu}
\address[author1]{College of Computer Science and Technology, Zhejiang University, Hangzhou 310027, China}
\address[author2]{Library and Information Center, Zhejiang University, Hangzhou 310027, China}

\begin{abstract}
Peer-to-Peer (P2P) systems have proved to be the most effective and popular file sharing applications in recent years. Previous studies mainly focus on the equal service and the differentiated service strategies when peers have no initial data before their download. In an upload-constrained P2P file sharing system, we model both the equal service process and the differentiated service process when peers' initial data distribution satisfies some special conditions, and also show how to minimize the time to get the file to any number of peers. The proposed models can reveal the intrinsic relations among the initial data amount, the size of peer set and the minimum last finish time. By using the models, we can also provide arbitrary degree of differentiated service to a certain number of peers. We believe that our analysis process and achieved theoretical results could provide fundamental insights into studies on bandwidth allocation and data scheduling, and can give helpful reference both for improving system performance and building effective incentive mechanism in P2P file sharing systems.
\end{abstract}
\end{premaker}

\section{Introduction}
Peer-to-Peer (P2P) systems have proved to be the most effective and popular file sharing application in recent years. BitTorrent (BT) [1] is the most popular P2P application due to its high degree of scalability and effective incentive mechanism to reduce free-riding.

Recently, most of studies on P2P systems are focused on social network, incentive mechanism [2][3], piece selection algorithm [4] (especially directed at media streaming) etc. However, the bandwidth allocation and data scheduling for differentiated service in application layer has been paid little attention. We consider this problem to be one of the most fundamental and important problems because it determines how the data and bandwidth resource is allocated among peers and plays a decisive role in respect of improving the performance of P2P systems. In addition, theoretical analysis for differentiated service can also give some insight into the design of incentive mechanism.

There have been a lot of analytical studies focused on the overlay topology, service capacity and piece scheduling etc., when peers are homogeneous with the same download and upload bandwidth. Yang et al. [5][6] propose a branching process for transient state and a Markov chain model for steady state of P2P systems. Qiu et al. [7] present a simple fluid model in BitTorrent-like system for the first time, and they evaluate the scalability, performance and efficiency of such kind of P2P systems. However, this model only focuses on the peer arrival rate, departure rate in steady state, and makes an unrealistic assumption that all peers have homogeneous upload bandwidth. Based on the fluid model in [7], Clevenot-Perronnin et al. [8] propose a two-set fluid model for service differentiation in BitTorrent-like content distribution systems, whereas they assume peers in the same class have the same bandwidth. Biersack et al. [9] propose three typical analytical models, i.e. linear model, $k$-tree model, $k$-Ptree model, to address the problem that how long it should take to distribute a file to $N$ peers. Results indicate that the service capacity of most P2P systems grows exponentially with the number of pieces in a file. Esposito et al. [10] adapt the fluid model in [7] to transient phase of P2P file disseminating and propose a seed scheduling algorithm called \emph{proportional fair seed scheduling} (PFS). They show that their solution can effectively reduce the average download time than the standard BitTorrent protocol.

Under a more general condition that peers have heterogeneous bandwidth, some studies investigate the system performance (mostly in terms of last finish time or average finish time) and fairness (mostly in terms of peers' contributions to the system) to provide fundamental principles and criteria for P2P protocol and application system design. Wu et al. [11] propose a \emph{centrally scheduled file distribution} (CSFD) protocol that can provably minimize the last finish time of a single source file distribution process. Mundinger et al. [12][13] characterize the problem that how to minimize the last finish time when peers have heterogeneous upload bandwidth and the file can only be divided into discrete pieces. They also evaluate the minimum last finish time when all peers receive equal service and the file can be divided into infinitely small pieces, and show how each peer should schedule its bandwidth and data to achieve the lower bound. Kumar et al. [14] investigate file distribution with differentiated service in two sets of peers, with the purpose of distributing the file as quickly as possible to all the first-set peers. Also, they derive explicit expressions for the minimum distribution time. They make an assumption that the second-set peers have no initial data before download. Mehyar et al. [15] extend the results in [14] and examine different performance criteria (e.g. last finish time, average finish time, etc.) by considering how the pieces and the bandwidth resources can be allocated in order to reach the theoretical bounds. Fan et al. [16] analyze the delicate tradeoff between system performance and system fairness in BitTorrent-like system, and show that the default configuration in BitTorrent puts more emphasis on fairness rather than performance. Chan et al. [17] deduce a theoretical bound on the minimum last finish time in terms of discrete cycles. They also evaluate several scheduling algorithm suitable for the push-based model, e.g., \emph{Rarest piece first}, \emph{Most Demanding node first} and the \emph{Maximum-flow} algorithm, and simulation results show that the \emph{Maximum-flow} algorithm outperforms all other algorithms. Ezovski et al. [18] show how each peer should use its upload bandwidth to minimize the average finish time in a fully-connected network without the presence of peer leaving by using the \emph{water-filling technique}. Wen et al. [19] extend Ezovski et al.'s studies by taking peer leaving condition into account. They present the \emph{reversed water filling algorithm} (RWF) in order to minimize the average finish time. However, no theoretical lower bound is calculated because of the complexity of the problem.

This paper is inspired by the seminal studies [12][14][15][18][19], and can be regarded as an extension of [14][15][18]. In addition, the analysis methods (e.g. the \emph{water-filling technique}) in [18][19] have some significant implications for our research. To the best of our knowledge, there is no theoretical analysis towards the file sharing in P2P network where peers with heterogeneous upload bandwidth already have some initial data before their download. In [12][15], the equal service model and the two-set differentiated service model (including the \emph{Multiplicity Theorem}) are first proposed, and the system performance is also evaluated in terms of the last finish time, etc. In [14], towards the multi-set problem, they consider a two-set system for simplicity, nevertheless, how the second-set peers can finish their download has not been further considered. Based on their work, we present extended equal service model and differentiated service model when peers' initial data distribution satisfies some special conditions (\emph{UCP condition}), deduce the \emph{$\varphi$-Multiplicity Theorem}, and show how to make the best use of all peers' upload bandwidth to minimize the last finish time for peers in the superior set. Both our analysis process and achieved theoretical results could provide fundamental insights into studies on bandwidth allocation and data scheduling, and can give helpful reference both for improving system performance and building effective incentive mechanism in P2P file sharing systems.

The rest of this paper is organized as follows. Section 2 sums up and refines a few fundamental results in previous work. Section 3 discusses the concept of multiplicity in detail and presents our extended equal service and differentiated service models. Some numerical results are given in section 4 and we make the conclusion in section 5.

\section{Preliminary}

\subsection{Assumptions and Metrics}

According to the actual application background and the need of the analysis, most of related works make the following reasonable assumptions, which will also be adopted in this paper:

\begin{itemize}
\item The file can be divided into infinitesimal pieces. Each peer can simultaneously upload to as many peers as needed. Although in real P2P application systems, the maximum concurrent uploads is a fixed value. This assumption is consistent with [12][13][14][18] etc.
\item Upload bandwidth rather than download bandwidth is the only bottleneck. This assumption holds because in reality users always have asymmetric access bandwidth. This assumption is consistent with [14][15][18][19] etc.
\end{itemize}

Unlike the P2P live streaming case where emphasis should be given onto the transmission delay [20] and the streaming capacity [21] etc., the following two metrics involving the download time should be considered in detail, as we assume that a file can only be used after its last bit is obtained by a peer.

\begin{itemize}
  \item \emph{Last finish time}: The time interval every peer in a specified peer set obtains the whole file.
  \item \emph{Bottleneck time}: The time interval that the source can distribute its fresh data (the data no one else has) to the swarm.
\end{itemize}

Obviously, for any set of peers, the last finish time is no less than the bottleneck time. One of the most important targets in this paper is to minimize the last finish time.

\subsection{Fundamental Results in Previous Work}

Given an upload-constrained fully connected P2P network, if all of $N$ peers receive the same service (\emph{equal service}) from a source, the minimum last finish time ${{T}_{L}}$ is given [12][14] by the following equation with symbols modified for consistency, where ${{C}_{0}}$ is the source's upload bandwidth, $U$ is the sum of all peers' upload bandwidth, and $F$ is the file size.

\begin{equation}\label{eq:1}
{{T}_{L}}=\frac{F}{\min \left\{ {{C}_{0}},\frac{{{C}_{0}}+U}{N} \right\}}
\end{equation}

If there are two sets of peers, and the first $L$ peers have superior priority (\emph{differentiated service}), the minimum last finish time of peers in the first set is given [14] by the following equation, where ${{C}_{N-L}}$ is the total upload bandwidth of the last $N-L$ peers.

\begin{equation}\label{eq:2}
{{T}_{L}}=\frac{F}{\min \left\{ {{C}_{0}},\frac{1}{L}\left( {{C}_{0}}+U-\frac{{{C}_{N-L}}}{L} \right) \right\}}
\end{equation}

The bottleneck time ${{t}_{b}}$ is calculated by its definition as:

\begin{equation}\label{eq:3}
{{t}_{b}}=\frac{F}{{{C}_{0}}}
\end{equation}

Furthermore, Mehyar et al. [15] propose the following theorem to calculate at most how many peers can finish their download in the bottleneck time.

\emph{Multiplicity Theorem}: It is possible to let the first $M$ peers finish their download in the bottleneck time ${{t}_{b}}$ if and only if:

\begin{equation}\label{eq:4}
{{C}_{0}}\le \sum\limits_{i=1}^{M}{\frac{{{c}_{i}}}{M-1}}+\sum\limits_{i=M+1}^{N}{\frac{{{c}_{i}}}{M}},
\end{equation}
where ${{c}_{i}}$ is the upload bandwidth of peer $i$. For convenience of expression, we use ${{C}_{M}}$ and ${{C}_{N-M}}$ to denote the total upload bandwidth of the first $L$ peers and the last $N-L$ peers respectively, hence above inequality can be rewritten as ${{C}_{0}}\le \frac{1}{M-1}{{C}_{M}}+\frac{1}{M}{{C}_{N-M}}$.

For a given P2P system, \emph{multiplicity} is defined as the largest $M$ such that inequality (4) holds.

It is worth noting that, the \emph{multiplicity} here refers specifically to the $M$ in the case that all peers are put in descending order by their upload bandwidth, as $M$ can reach the maximum in this case. In this paper, the \emph{multiplicity} we refer to is the largest integer $M$ such that inequality (4) holds, where peers are put in random order in advance. The discussions in what follows will not put a limit on the order of peers either.

Owing to space constraints, the strategies to achieve the above theoretical lower (upper) bounds in related literatures will not be given in this paper.

\subsection{Differentiated service model when peers have no initial data}

As section 2.2 shows, ${{T}_{L}}$ has already been deduced, however, there is no unified form for ${{T}_{L}}$ in the differentiated service model [14] and the \emph{Multiplicity Theorem} [15]. In this section, we will reformulate ${{T}_{L}}$ when there is no initial data before download.

For convenience of expression, we define the \emph{multiplicity function}:

\begin{equation}\label{eq:5}
F(M)=\frac{1}{M-1}{{C}_{M}}+\frac{1}{M}{{C}_{N-M}}
\end{equation}

When $M=N$, $F(M)$ is defined by $\frac{1}{M-1}{{C}_{M}}$; when $M=1$, $F(M)$ becomes infinity [15].

According to the magnitude of $L$, there are two cases to be considered.

\noindent1)	$L\le M$

In this case, the last finish time of the first $L$ peers is equivalent to the bottleneck time, i.e.:

\begin{equation}\label{eq:6}
{{T}_{L}}={{t}_{b}}
\end{equation}

\noindent2)	$L>M$

Since $L>M$, we can deduce $F(L)<F(M)$. In this case, the first $L$ peers can reach the maximum download rate, and ${{T}_{L}}$ can be calculated as follows:

\begin{equation}\label{eq:7}
{{T}_{L}}=\frac{F}{\frac{1}{L}({{C}_{0}}+U-\frac{1}{L}{{C}_{N-L}})}
\end{equation}

\section{Extended equal service and differentiated service models}

\subsection{Data Distribution}

Researches in [12][13][14][15] etc. all make an assumption that each peer has no data before download. In this paper, in order to study how the initial data will put an effect on the whole file dissemination process, we will consider the cases when the data distribution satisfies two special conditions.

According to the previous section, in order to provide peers in $L1$ with superior service, peer $L+i$ in $L2$ has to download at certain rate $\mu \le \frac{1}{L}{{c}_{L+i}}$ from the source, then replicates this data stream to all $L$ peers in $L1$. In the case of $L>M$, when all $L$ peers finish downloading in ${{T}_{L}}$, the data amount peer $L+i$ has is $\frac{1}{L}{{c}_{L+i}}$ (refer to Fig.1). As a consequence, any peer in $L2$ has entirely different data proportional to its upload bandwidth from other peers in $L2$ after the download process. Obviously, the initial data that peers in $L2$ already have after the current download process will directly influence how these peers can finish their download in further download process. In the remaining sections, we call these data the \emph{forwarding data}. Now we introduce the \emph{UP condition} as follows:

\begin{figure}[!h]
\centering
\includegraphics[angle=0, width=0.45\textwidth]{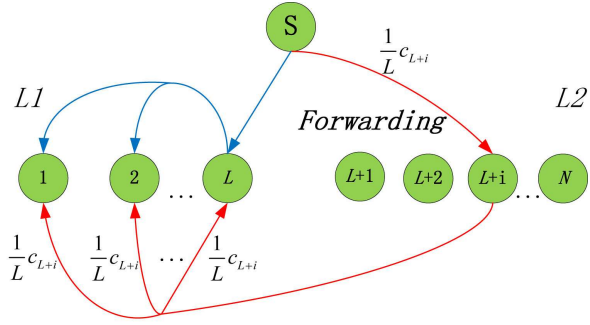}
\caption{Illustration of the \emph{UP condition}}
\label{forwarding}
\end{figure}

Unique and Proportional condition: (\emph{UP condition}) The data contained within the same peer set is a subset of the whole file, and it has the following two characteristics:
\begin{itemize}
\item Each file piece must be owned by only one peer (\emph{unique data});
\item The ratio of the unique data amount to the upload bandwidth of each peer is equal.
\end{itemize}

Let ${{a}_{i}}$ denote the data peer $i$ owns before download, and the ratio of all peers' data amount to the file size $F$ is denoted by $\varphi$ (when $\varphi$ is 0, it turns to the no-data case), i.e.:

\begin{equation}\label{eq:8}
\varphi =\sum\limits_{1}^{N}{{{a}_{i}}}/F
\end{equation}

We can formally describe the \emph{UP condition} as follows:

\begin{equation}\label{eq:9}
	\left\{ \begin{array}{l}
   {{t}_{a}}:=\frac{{{a}_{1}}}{{{c}_{1}}}=\frac{{{a}_{2}}}{{{c}_{2}}}=...=\frac{{{a}_{N}}}{{{c}_{N}}}=\frac{\varphi F}{U}  \\
   {{a}_{i}}\cap {{a}_{j}}=\varnothing ,\ \forall i\ne j  \\
   \bigcup\limits_{i=1}^{N}{{{a}_{i}}}\subseteq F  \\
\end{array} \right.,
\end{equation}
where ${{t}_{a}}$ is defined as the time interval that each peer can distribute its own unique data to other peers. According to Equation (9) and the definition of $\varphi$ , we have $\varphi \le 1$.

As a supplement to the \emph{UP condition}, we also introduce the \emph{UCP condition} as follows:

Unique-Common and Proportional condition: (\emph{UCP condition}) The data contained within the same peer set is a subset of the whole file, and it has the following two characteristics:

\begin{itemize}
\item Each file piece must be owned by only one peer (\emph{unique data}) or be owned by all peers (\emph{common data});
\item The ratio of the unique data amount to the upload bandwidth of each peer is equal.
\end{itemize}

\begin{figure}[!h]
\centering
\includegraphics[angle=0, width=0.3\textwidth]{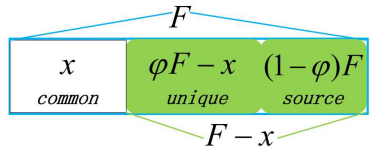}
\caption{Data distribution (\emph{UP condition} and \emph{UCP condition})}
\label{fig:fi_F}
\end{figure}

Assume the common data, the unique data and the data owned only by the source be $x$, $\varphi F-x$ and $(1-\varphi )F$ respectively, as Fig. 2 shows. The \emph{UCP condition} for a file $F$ is equivalent to the \emph{UP condition} for a file $F-x$, hence, without loss of generality, we only consider the \emph{UP condition} in our extended models.

The data amount owned by all peers is denoted by ${{F}_{a}}$, and the data amount owned only by the source is denoted by ${{F}_{0}}$, then:

\begin{equation}\label{eq:10}
{{F}_{a}}:=\sum\limits_{1}^{N}{{{a}_{i}}}=\varphi F, {{F}_{0}}:=(1-\varphi )F
\end{equation}

In order to download the whole file, the source has to upload data ${{F}_{0}}$ to other peers. Let ${{t}_{0}}$ be the bottleneck time, then:

\begin{equation}\label{eq:11}
{{t}_{0}}=\frac{{{F}_{0}}}{{{C}_{0}}}
\end{equation}

${{t}_{0}}$ must be the tightly lower bound of ${{T}_{L}}$, therefore we have:

\begin{equation}\label{eq:12}
{{T}_{L}}\ge {{t}_{0}}
\end{equation}

If ${{t}_{a}}={{t}_{0}}$, $\varphi$ can be expressed as $\frac{U}{{{C}_{0}}+U}$, and this value is denoted by ${{\varphi }_{0}}$ for convenience, as equation (13) shows. Therefore $\varphi \ge {{\varphi }_{0}}$ is equavalent to ${{t}_{0}}\le {{t}_{a}}$, and vice versa.

\begin{equation}\label{eq:13}
{{\varphi }_{0}}:=\frac{U}{{{C}_{0}}+U}
\end{equation}

\subsection{Equal Service}

In this section, we will extend the single-set problem mentioned in [14] when peers' initial data distribution satisfies the \emph{UP condition}, and calculate ${{T}_{L}}$ according to the relationship between $\varphi$ and ${{\varphi }_{0}}$, there are two cases to be considered.

\noindent1) $\varphi \ge {{\varphi }_{0}}$ (${{t}_{0}}\le {{t}_{a}}$)

Without $N$ peers' own upload bandwidth, it takes the source $N{{t}_{0}}$ to send ${{F}_{0}}$ to all peers, and without the source's help, it takes all $N$ peers $(N-1){{t}_{a}}$ to get ${{F}_{a}}$. According to the relationship of $N{{t}_{0}}$ and $(N-1){{t}_{a}}$, we have the following two cases.

a) $(N-1){{t}_{a}}\ge N{{t}_{0}}$

In this case, the source should divide part of its bandwidth $x$ to help peers downloading ${{F}_{a}}$, thus we have (refer to Fig. 3(a)):

\begin{equation}\label{eq:14}
\frac{{{F}_{0}}}{\frac{1}{N}({{C}_{0}}-x)}=\frac{{{F}_{a}}}{\frac{1}{N-1}(U+x)}={{T}_{L}}
\end{equation}

When $\varphi =1$, peers have the entire file $F$. The source peer can help every peer send its own data to other $N-1$ peers, and ${{T}_{L}}$ can be directly expressed as $\frac{(N-1)F}{{{C}_{0}}+U}$.

b) $(N-1){{t}_{a}}<N{{t}_{0}}$

In this case, we have ${{t}_{0}}\le (N-1){{t}_{a}}<N{{t}_{0}}$. We can find that by taking some bandwidth $x$ out of $U$ to download ${{F}_{0}}$, it will take the same amount of time to download ${{F}_{0}}$ and ${{F}_{a}}$ (refer to Fig. 3(b)), then we have:

\begin{equation}\label{eq:15}
\frac{{{F}_{0}}}{\frac{1}{N}({{C}_{0}}+x)}=\frac{{{F}_{a}}}{\frac{1}{N-1}(U-x)}={{T}_{L}}
\end{equation}

Equations (14) and (15) have the same solution:

\begin{equation}\label{eq:16}
{{T}_{L}}=\frac{(N-\varphi )F}{{{C}_{0}}+U}
\end{equation}

As all upload bandwidth can be utilized, ${{T}_{L}}$ can also be viewed as the ratio of the total data required to transmit to the total upload bandwidth. Obviously, equation (16) still holds when $\varphi =1$.

\noindent2) $\varphi <{{\varphi }_{0}}$ (${{t}_{0}}>{{t}_{a}}$)

a) $(N-1){{t}_{a}}\ge {{t}_{0}}$

In this case, we have $N{{t}_{0}}>(N-1){{t}_{a}}\ge {{t}_{0}}$, part of $U$ is used to exchange ${{F}_{0}}$, the bandwidth can also be allocated as Fig. 3(b) shows, and ${{T}_{L}}$ can also be calculated by equation (16).

b) $(N-1){{t}_{a}}<{{t}_{0}}$

Let ${{U}_{N}}$ be the total upload bandwidth needed to distribute ${{F}_{a}}$ to all peers, then the remaining upload bandwidth ${{U}_{r,N}}$ is $U-{{U}_{N}}$, i.e.:

\begin{equation}\label{eq:17}
{{U}_{N}}=\frac{(N-1){{F}_{a}}}{{{t}_{0}}}, {{U}_{r,N}}=U-{{U}_{N}}
\end{equation}

As peers must download ${{F}_{0}}$ from the source, the download speed of ${{F}_{0}}$ depends on the relation between ${{U}_{r,N}}$ and ${{C}_{0}}$:

i) If ${{U}_{r,N}}\ge (N-1){{C}_{0}}$, ${{U}_{r,N}}$ is large enough to ensure the download speed of ${{F}_{0}}$ to be ${{C}_{0}}$. As a result, the last finish time is constraint by the bottleneck time, i.e. ${{T}_{L}}={{t}_{0}}$.

ii) If ${{U}_{r,N}}<(N-1){{C}_{0}}$, ${{U}_{r,N}}$ cannot maximize the download speed of ${{F}_{0}}$. We can also use the bandwidth allocation strategy in equation (15) (refer to Fig. 3(b)), and ${{T}_{L}}$ can also be calculated by equation (16).

\begin{figure}[h]
\centering
\subfigure[]{
\label{fig:subfig:a}
\includegraphics[width=0.22\textwidth]{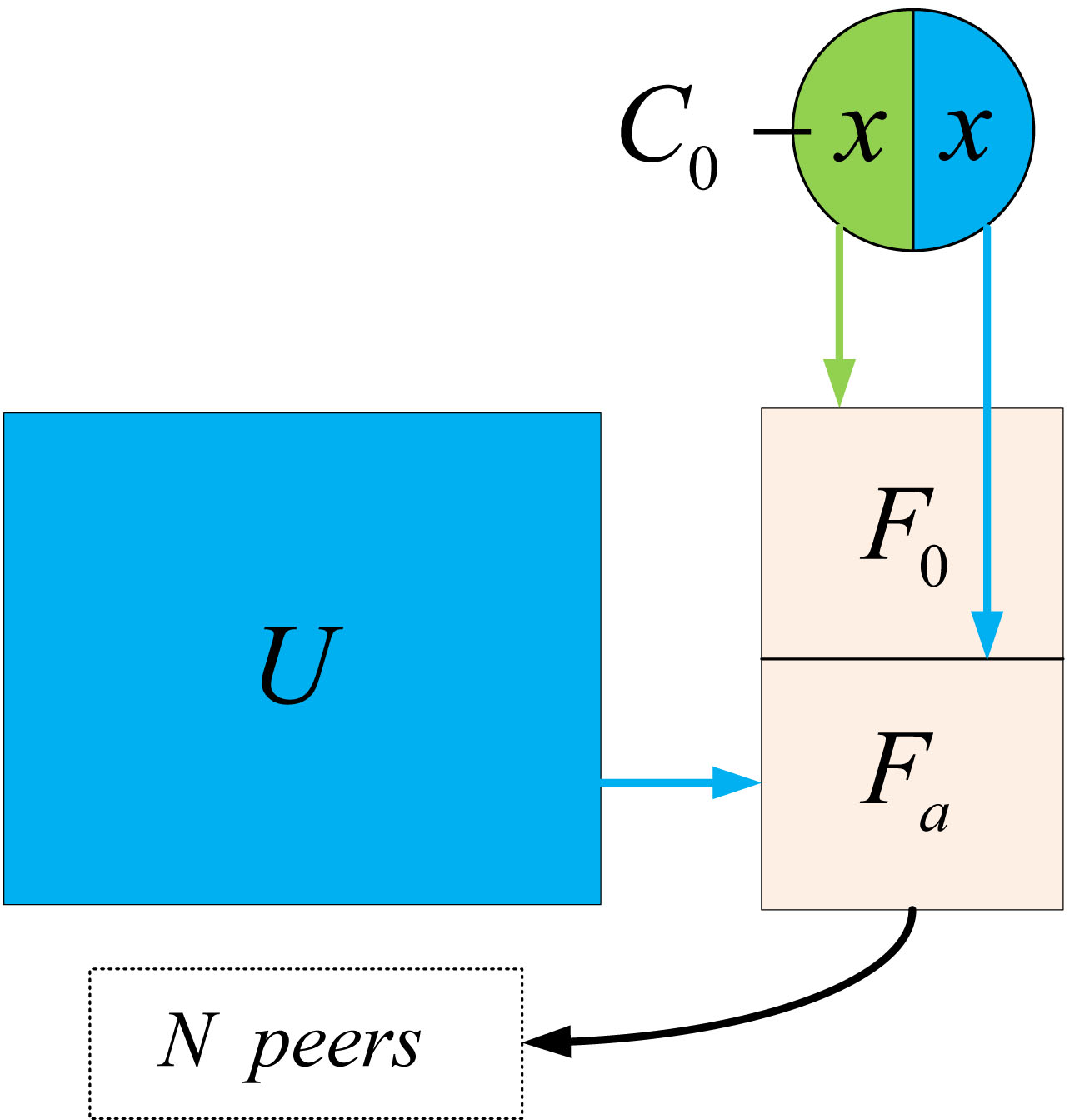}}
\hspace{1in}
\subfigure[]{
\label{fig:subfig:b}
\includegraphics[width=0.22\textwidth]{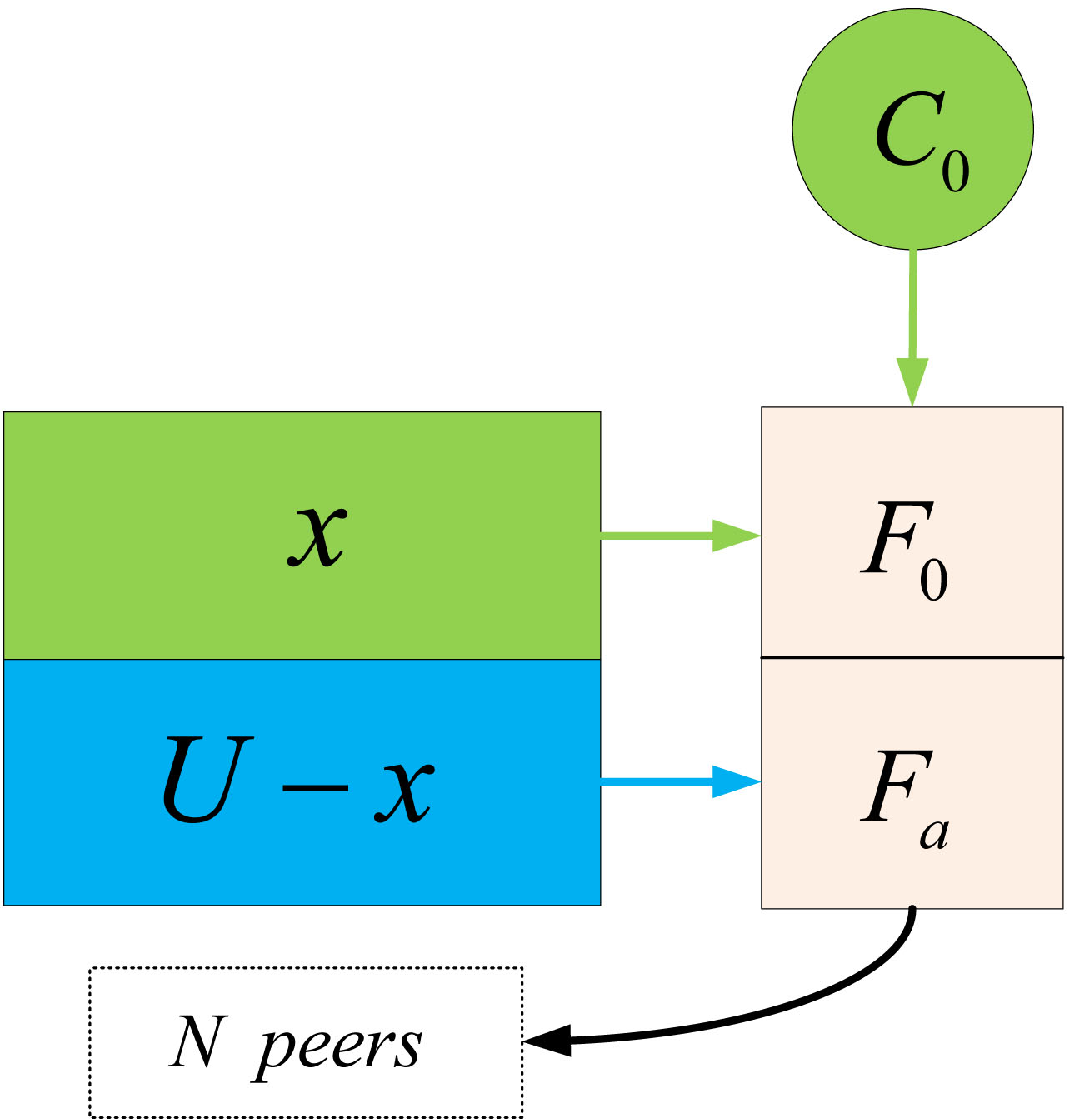}}
\caption{Bandwidth allocation strategies (\emph{equal service})}
\label{fig:pc}
\end{figure}

\subsection{Differentiated Service}

In this section, the multi-set problem mentioned in [14] and the \emph{Multiplicity Theorem} mentioned in [15] will be extended and generalized under the \emph{UP condition}.

\noindent1) $\varphi \ge {{\varphi }_{0}}$ (${{t}_{0}}\le {{t}_{a}}$)

If $\varphi >{{\varphi }_{0}}$, no one can finish its download in ${{t}_{0}}$, i.e. $M=0$, while in the case of $\varphi ={{\varphi }_{0}}$, there is one and only one can complete in ${{t}_{0}}$, i.e. $M=1$.

a) $L=1$

In this case, the source and all peers from the second peer upload to the first peer at full speed, the minimum last finish time can be calculated as:

\begin{equation}\label{eq:18}
{{T}_{L}}=\frac{F-{{a}_{1}}}{{{C}_{0}}+U-{{c}_{1}}}
\end{equation}

b) $L>1$

Denote the data amount of the first $L$ peers and the last $N-L$ peers by $F_L$ and $F_{N-L}$ respectively. If ${{C}_{N-L}}$ is fully used for transmitting ${{F}_{N-L}}$ to the first $L$ peers, the time interval is $L{{t}_{a}}$, then, in order to exchange ${{F}_{L}}$ in $L{{t}_{a}}$, $\frac{L-1}{L}{{C}_{L}}$ of their total upload bandwidth is used and $\frac{1}{L}{{C}_{L}}$ is left.

Without the source's help, if all of $\frac{1}{L}{{C}_{L}}$ is still used to download ${{F}_{a}}$, suppose it takes the first $L$ peers ${{T}_{L}}'$ to obtain ${{F}_{a}}$, we have ${{t}_{a}}\le (L-1){{t}_{a}}<{{T}_{L}}'<L{{t}_{a}}$, which means that all of $\frac{1}{L}{{C}_{L}}$ can be fully utilized. Suppose the first $L$ peers divide part of their bandwidth $x$ to help the last $N-L$ peers downloading ${F}_{N-L}$ (refer to Fig. 4), then we have:

\begin{equation}\label{eq:19}
\frac{{{F}_{L}}}{\frac{1}{L-1}\left( {{C}_{L}}-x \right)}=\frac{{{F}_{N-L}}}{\frac{1}{L}\left( {{C}_{N-L}}+x \right)}={{T}_{L}}'
\end{equation}

Solving above equation, we have:

\begin{equation}\label{eq:20}
{{T}_{L}}'=\frac{L{{F}_{a}}-{{F}_{L}}}{U}
\end{equation}

\begin{figure}[!h]
\centering
\includegraphics[angle=0, width=0.4\textwidth]{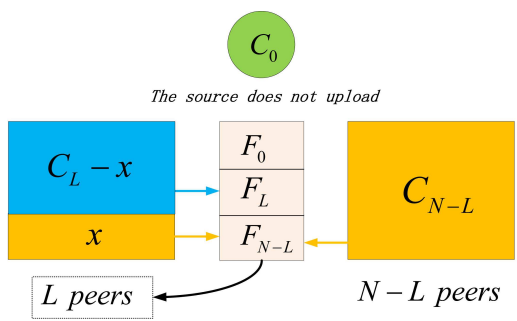}
\caption{Bandwidth allocation strategy without the source's help (\emph{differentiated service})}
\label{fig:levfig}
\end{figure}

Without using $\frac{1}{L}{{C}_{L}}$, it takes the source at most $L{{t}_{0}}$ to send ${{F}_{0}}$ to the first $L$ peers. The most important question of the bandwidth allocation strategy is how $\frac{1}{L}{{C}_{L}}$ should be allocated.

If the first $L$ peers can fully exchange their own data ${{F}_{L}}$ within $T$, the upload bandwidth required is denoted by ${{U}_{L}}(L,T)$. Similarly, the upload bandwidth required to make the last $N-L$ peers upload their own data ${{F}_{N-L}}$ to the first $L$ peers within $T$ is denoted by ${{U}_{N-L}}(L,T)$. Therefore we have:

\begin{equation}\label{eq:21}
{{U}_{L}}(L,T)=\frac{(L-1){{F}_{L}}}{T}, {{U}_{N-L}}(L,T)=\frac{L{{F}_{N-L}}}{T}
\end{equation}

According to the relationship between $L{{t}_{0}}$ and ${{T}_{L}}'$, we have the following two cases.

i) $L{{t}_{0}}<{{T}_{L}}'$

If $\varphi$ is larger than some critical value, the source should split its upload bandwidth into three parts ${{C}_{0}}-{{x}_{L}}-{{x}_{N-L}}$, ${{x}_{L}}$ and ${{x}_{N-L}}$ to help downloading ${{F}_{0}}$, ${{F}_{L}}$ and ${{F}_{N-L}}$ respectively in the same time ${{T}_{L}}$, as Fig. 5(a) shows. Then we have:

\begin{equation}\label{eq:22}
\frac{{{F}_{0}}}{\frac{1}{L}\left( {{C}_{0}}-{{x}_{L}}-{{x}_{N-L}} \right)}={{T}_{L}}
\end{equation}

\begin{equation}\label{eq:23}
\left( {{U}_{L}}(L,{{T}_{L}})-{{x}_{L}} \right)+\left( {{U}_{N-L}}(L,{{T}_{L}})-{{x}_{N-L}}-{{C}_{N-L}} \right)={{C}_{L}}
\end{equation}

Combining equation (22) and (23), we obtain:

\begin{equation}\label{eq:24}
{{T}_{L}}=\frac{LF-{{F}_{L}}}{{{C}_{0}}+U}
\end{equation}

When $\varphi =1$, peers have the entire file $F$, then ${{t}_{0}}=0$. All of the source's upload bandwidth is used to transmit ${{F}_{L}}$ and ${{F}_{N-L}}$, then we have:

\begin{equation}\label{eq:25}
{{x}_{L}}+{{x}_{N-L}}={{C}_{0}}, {{F}_{L}}+{{F}_{N-L}}=F
\end{equation}

Combining equation (25) and (23), we get the same solution for ${{T}_{L}}$, as equation (24) shows.

ii) $L{{t}_{0}}\ge {{T}_{L}}'$

In this case, part of ${{C}_{L}}$ should be used to download ${{F}_{0}}$, and we denote it by ${{U}_{r,L,N-L}}(L,T)$, then:

\begin{equation}\label{eq:26}
{{U}_{r,L,N-L}}(L,T)=U-{{U}_{L}}(L,T)-{{U}_{N-L}}(L,T)
\end{equation}

In order to make ${{F}_{L}}$, ${{F}_{N-L}}$ and ${{F}_{0}}$ finished in the same time ${{T}_{L}}$, the upload bandwidth can be allocated as Fig. 5(b) shows, and the following equation must hold:

\begin{equation}\label{eq:27}
\frac{{{F}_{0}}}{\frac{1}{L}\left[ {{C}_{0}}+{{U}_{r,L,N-L}}(L,{{T}_{L}}) \right]}={{T}_{L}}
\end{equation}

Solving above equation, ${{T}_{L}}$ is also expressed as equation (24).

Notice that, all above bandwidth allocation strategies in i) and ii) do not need the last $N-L$ peers to forward any part of ${{F}_{0}}$ from the source.

\noindent2) $\varphi <{{\varphi }_{0}}$ (${{t}_{0}}>{{t}_{a}}$)

Suppose $L'{{t}_{a}}<{{t}_{0}}\le (L'+1){{t}_{a}}$ ($L'\in [1,N-1)$) or $(L'-1){{t}_{a}}<{{t}_{0}}$ ($L'=N-1$), where $L'$ is an integer. For convenience of expression, we use $L'{{t}_{a}}<{{t}_{0}}\le (L'+1){{t}_{a}}$ ($L'\in [1,N-1]$) to indicate the above two cases in what follows, it has no impact on the correctness of the analysis. There are three cases to be considered according to the relationship between $L$ and $L'+1$.

a) $L=L'+1$

In this case, ${{T}_{L}}'$ may be equal to, larger or smaller than ${{t}_{0}}$.

i) ${{T}_{L}}'>{{t}_{0}}$

We define ${{T}_{0}}$ as the minimum downloading time of ${{F}_{0}}$ by using $\frac{1}{L}{{C}_{L}}$, i.e.:

\begin{equation}\label{eq:28}
{{T}_{0}}:=\max (\frac{{{F}_{0}}}{\frac{1}{L}\left( {{C}_{0}}+\frac{1}{L}{{C}_{L}} \right)},{{t}_{0}})
\end{equation}

Case A: ${{T}_{0}}>L{{t}_{a}}$. In this case, as can be illustrated by Fig. 5(c), part of upload bandwidth of the last $N-L$ peers should be used for forwarding ${{F}_{0}}$ from source to the first $L$ peers, which is denoted by ${{U}_{r,N-L}}(L,T)$. Correspondingly, part of uplaod bandwidth of the first $L$ peers should also be used for downloading ${{F}_{0}}$, which is denoted by ${{U}_{r,L}}(L,T)$. Then we have:

\begin{equation}\label{eq:29}
{{U}_{r,L}}(L,T)={{C}_{L}}-{{U}_{L}}(L,T), {{U}_{r,N-L}}(L,T)={{C}_{N-L}}-{{U}_{N-L}}(L,T)
\end{equation}

In order to make ${{F}_{0}}$ finished in ${{T}_{L}}$, we have:

\begin{equation}\label{eq:30}
\frac{{{F}_{0}}}{\frac{1}{L}\left[ {{C}_{0}}+(1-\frac{1}{L}){{U}_{r,N-L}}(L,{{T}_{L}})+{{U}_{r,L}}(L,{{T}_{L}}) \right]}={{T}_{L}}
\end{equation}

Notice that, ${{U}_{r,N-L}}(L,{{T}_{L}})$ must be multiplied by a factor $1-\frac{1}{L}$, as the source must provide data $\frac{1}{L}{{U}_{r,N-L}}(L,{{t}_{0}})$ to the last $N-L$ peers. This can be illustrated by Fig. 1. Most of all, the data distribution of the last $N-L$ peers will satisfy the \emph{UCP condition} as well after the first $L$ peers finish their download.

Solving equation (30), we get:

\begin{equation}\label{eq:31}
{{T}_{L}}=\frac{(L-\varphi )F}{{{C}_{0}}+U-\frac{1}{L}{{C}_{N-L}}}
\end{equation}

Case B: ${{T}_{0}}\le L{{t}_{a}}$. In this case, $\frac{1}{L}{{C}_{L}}$ can be fully used for downloading ${{F}_{0}}$ in a time interval smaller than $L{{t}_{a}}$, then we can use the same bandwidth allocation strategy as Fig. 5(b) shows to make ${{F}_{0}}$ and ${{F}_{a}}$ finished in the same time ${{T}_{L}}$, and ${{T}_{L}}$ can be also expressed as equation (24).

ii) ${{T}_{L}}'\le {{t}_{0}}$

Case A: ${{T}_{0}}>L{{t}_{a}}$. This case is similar to case A of i), and ${{T}_{L}}$ can be calculated in the same way (refer to Fig. 4 (c)), as equation (31) shows.

Case B: ${{T}_{0}}\le L{{t}_{a}}$. If ${{t}_{0}}<(L'+1){{t}_{a}}$, ${{F}_{L}}$ and ${{F}_{N-L}}$ cannot be finished in ${{t}_{0}}$ without using $\frac{1}{L}{{C}_{L}}$. From equations (21) and (26), we can calculate ${{U}_{r,L,N-L}}(L,{{t}_{0}})=U-\frac{(L-1)F+{{F}_{N-L}}}{{{t}_{0}}}$. In order to make the first $L$ peers finish downloading ${{F}_{0}}$ within ${{t}_{0}}$, we can allocate all upload bandwidth as Fig. 5(b) shows, and the following inequality must hold:

\begin{equation}\label{eq:32}
{{U}_{r,L,N-L}}(L,{{t}_{0}})\ge (L-1){{C}_{0}}
\end{equation}

Otherwise, the last finish time cannot reach ${{t}_{0}}$, and ${{T}_{L}}$ can be expressed as equation (24).

b) $L<L'+1$

In this case, we have ${{T}_{L}}'<L'{{t}_{a}}<{{t}_{0}}$. Fig. 5(c) illustrates the bandwidth allocation strategy. In order to make the first $L$ peers finish download within ${{t}_{0}}$, the following inequality must hold:

\begin{equation}\label{eq:33}
{{C}_{0}}\le \frac{1}{L}\left[ {{C}_{0}}+(1-\frac{1}{L}){{U}_{r,N-L}}(L,{{t}_{0}})+{{U}_{r,L}}(L,{{t}_{0}}) \right]
\end{equation}

Substituting equation (29) into inequality (33), we can deduce the following inequality:

\begin{equation}\label{eq:34}
{{C}_{0}}\le (1-\varphi )F(L)
\end{equation}

c) $L>L'+1$

In this case we have ${{T}_{L}}'>(L'+1){{t}_{a}}>{{t}_{0}}$, therefore the discussions and equations to calculate ${T_L}$ are the same as in case i) of a).

\begin{figure}[h]
\centering
\subfigure[]{
\label{fig:subfig:a}
\includegraphics[width=0.32\textwidth]{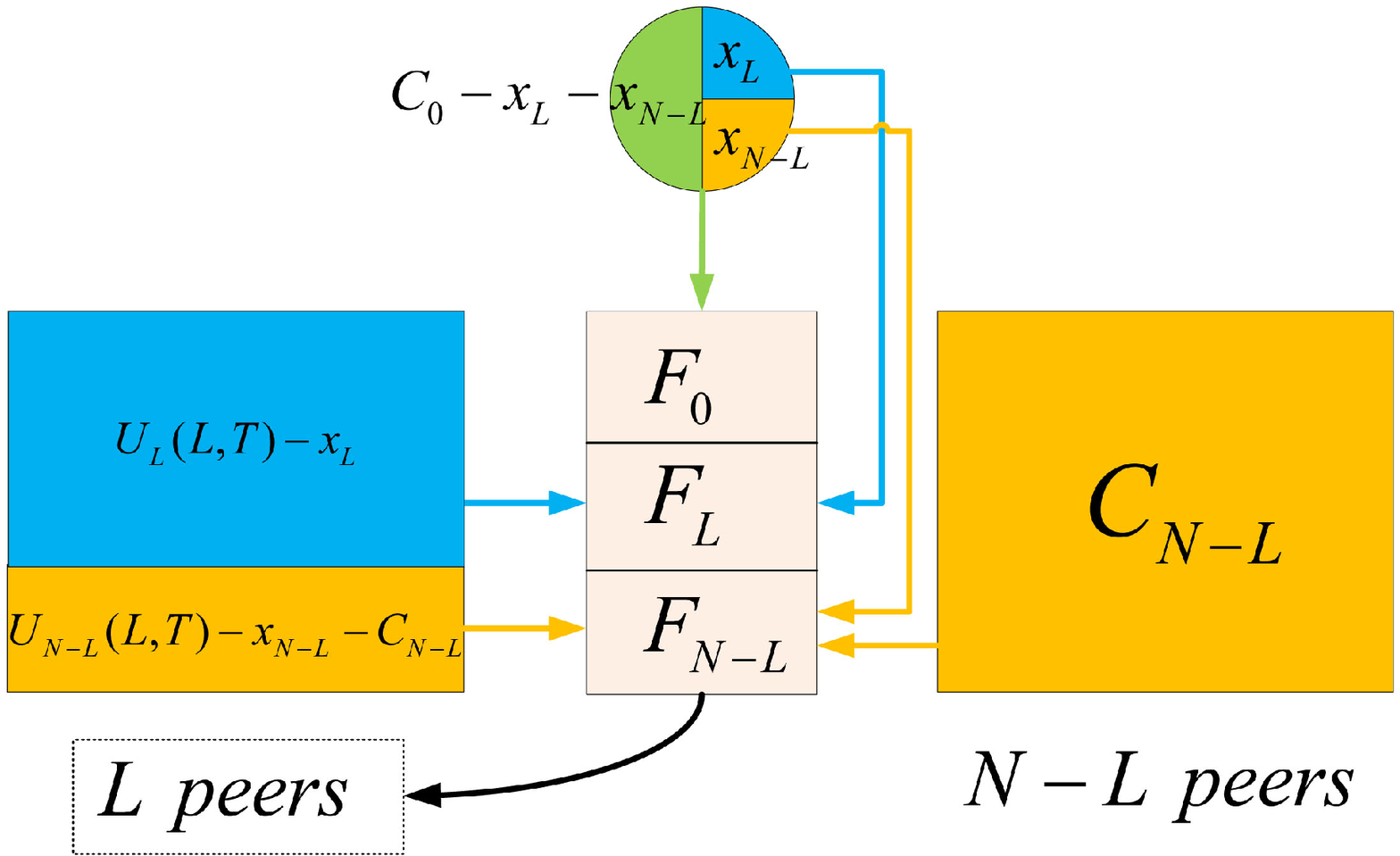}}
\subfigure[]{
\label{fig:subfig:b}
\includegraphics[width=0.32\textwidth]{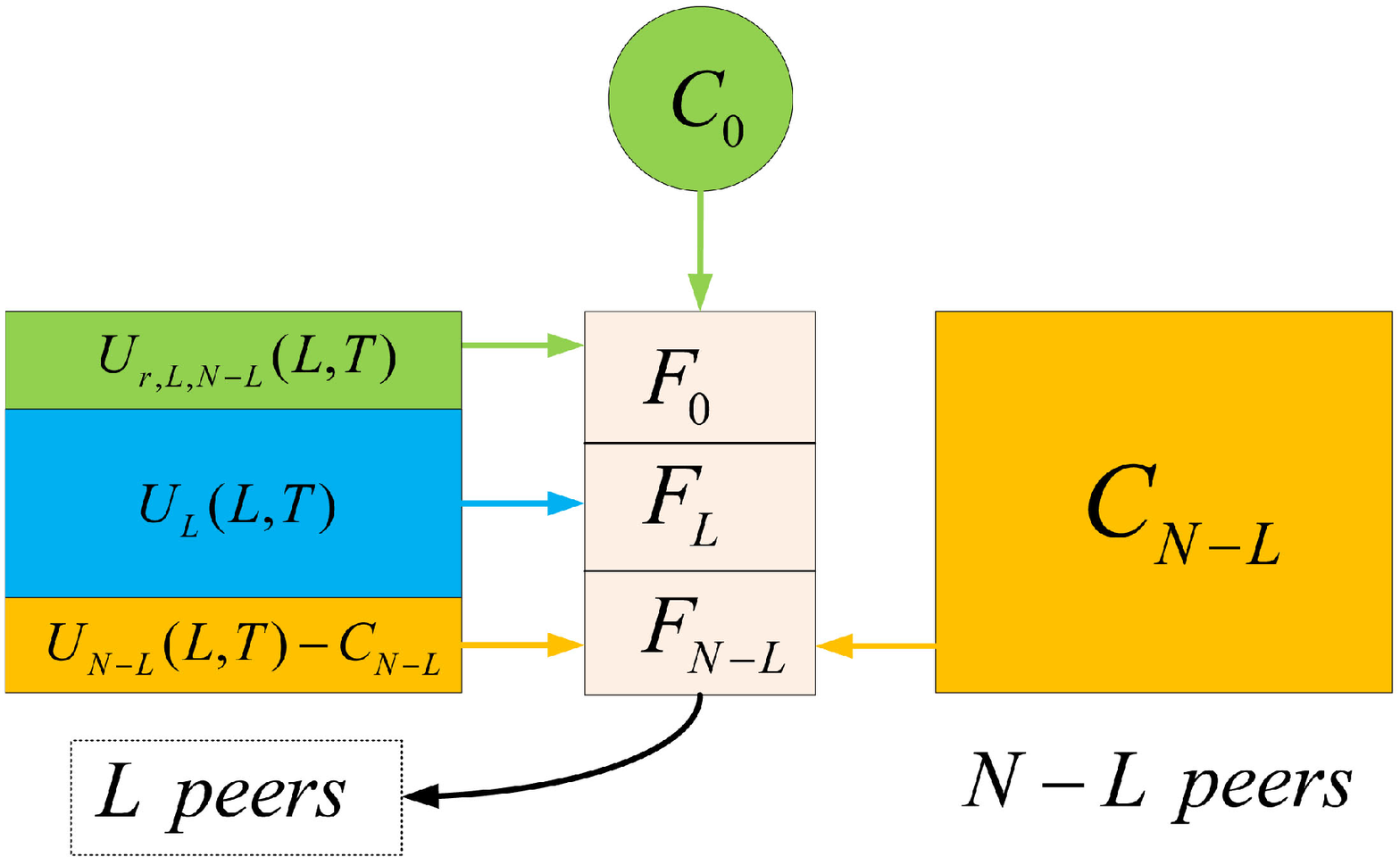}}
\subfigure[]{
\label{fig:subfig:c}
\includegraphics[width=0.32\textwidth]{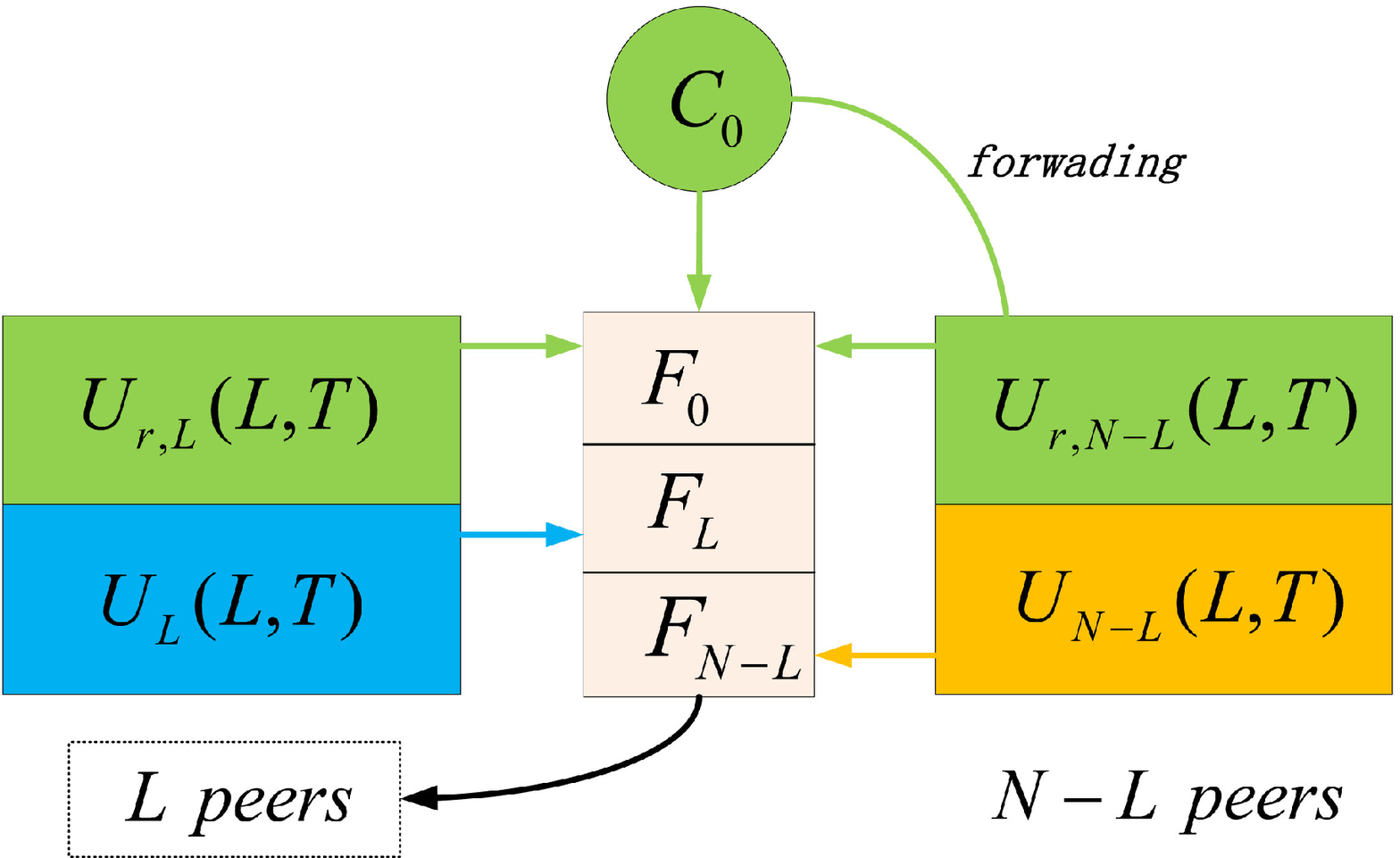}}
\caption{Bandwidth allocation strategies (\emph{differentiated service}). In (a) and (b), the last $N-L$ peers need not to forward ${{F}_{0}}$ from the source to the first $L$ peers, while the opposite is true in (c).}
\label{fig:subfig}
\end{figure}

\textbf{Theorem 1}: Suppose $L'{{t}_{a}}<{{t}_{0}}\le (L'+1){{t}_{a}}$ ($L'\in [1,N-1]$), if inequality (34) is true when $L=l+1$, where $l+1\le L'$, it is also true when $L\le l$; if inequality (32) is true when $L=L'+1$, inequality (34) is true when $L\le L'$.

\emph{Proof:} The proof of Theorem 1 is in the appendix.

Based on the above conclusions, we can generalize the \emph{Multiplicity Theorem} [15] to the following \emph{$\varphi$-Multiplicity Theorem}.

\textbf{Theorem 2} (\emph{$\varphi$-Multiplicity Theorem}): For a given P2P system, if the initial data distribution satisfies the \emph{UCP condition} and the initial data ratio is $\varphi$, it is possible to let the first $M$ peers finish their download in the bottleneck time ${t_0}$. We define the largest integer $M$ as the \emph{$\varphi$-Multiplicity}, and it can be calculated as follows:

\begin{itemize}
\item If ${{t}_{0}}<{{t}_{a}}$, no peer can finish its download in ${{t}_{0}}$, then $M=0$.
\item If ${{t}_{0}}={{t}_{a}}$, there is one and only one peer can finish its download in ${{t}_{0}}$, then $M=1$.
\item If $L'{{t}_{a}}<{{t}_{0}}\le (L'+1){{t}_{a}}$ ($L'\in [1,N-1]$), there are two cases: If inequality (32) is true when $L=L'+1$, we have $M=L'+1$; otherwise, $M$ equals to the largest integer $L$ no more than $L'$ such that inequality (34) is true.
\end{itemize}

Fig. 6 illustrates the \emph{$\varphi$-Multiplicity Theorem} in terms of the relation between ${t_0}$ and ${t_a}$ (or the variation of $\varphi$).

\begin{figure}[!h]
\centering
\includegraphics[angle=0, width=0.60\textwidth]{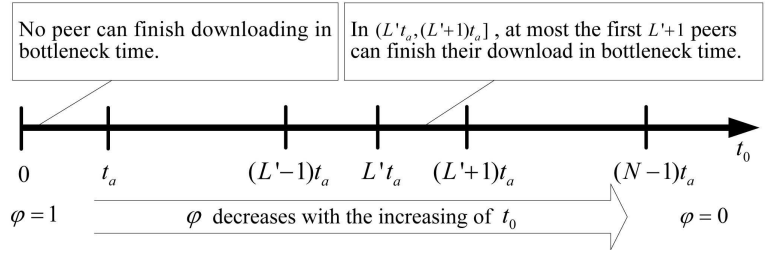}
\caption{Illustration of the \emph{$\varphi$-Multiplicity Theorem}}
\label{fig:levfig}
\end{figure}

Similar to section 2.3, we can calculate ${T_L}$ as follows according to the \emph{$\varphi$-Multiplicity Theorem}.

\noindent1) $L\le M$

In this case, ${T_L}$ must be constrained by the bottleneck time ${t_0}$ , i.e. ${{T}_{L}}={{t}_{0}}$.

\noindent2) $L>M$

a) If ${{T}_{0}}\le L{{t}_{a}}$, ${T_L}$ is calculated by equation (24).

b) If ${{T}_{0}}>L{{t}_{a}}$, ${T_L}$ is calculated by equation (31).

\section{Numerical results}
Suppose there are 18 peers with upload bandwidth 10, 10, 9, 9, 8, 8, 7, 7, 6, 6, 5, 5, 4, 4, 3, 3, 2, 2 (KB) respectively and one source peer distributing a file with the size of 1000KB. Fig. 7 gives the numerical results of both the original equal service and differentiated service models (i.e. $\varphi=0$) and our extended models (i.e. $\varphi \in (0,1]$).

We examine ${T_L}$ under $\varphi$ and $L$ in these models. Fig. 7 shows that, the magnitude of \emph{$\varphi$-Multiplicity} decreases with the increasing of $\varphi$. For a given $\varphi$, ${T_L}$ remains the same when $L$ is smaller or equal to the \emph{$\varphi$-Multiplicity}, and increases rapidly with the increasing of $L$ when $L$ is larger than the \emph{$\varphi$-Multiplicity}.

It is worthwhile to note that a smaller $\varphi$ will seriously limit the utilization of peers' upload bandwidth, especially when $L$ is small, hence will lead to a larger ${T_L}$. This is because when $\varphi$ is small, it will take the source peer a long time to distribute its unique data $F_0$ to other peers. On the other hand, when $L$ is large, all of peers' upload bandwidth will be better utilized, the increasing of $\varphi$ will have slight influence on ${T_L}$.

\begin{figure}[!h]
\centering
\includegraphics[angle=0, width=0.6\textwidth]{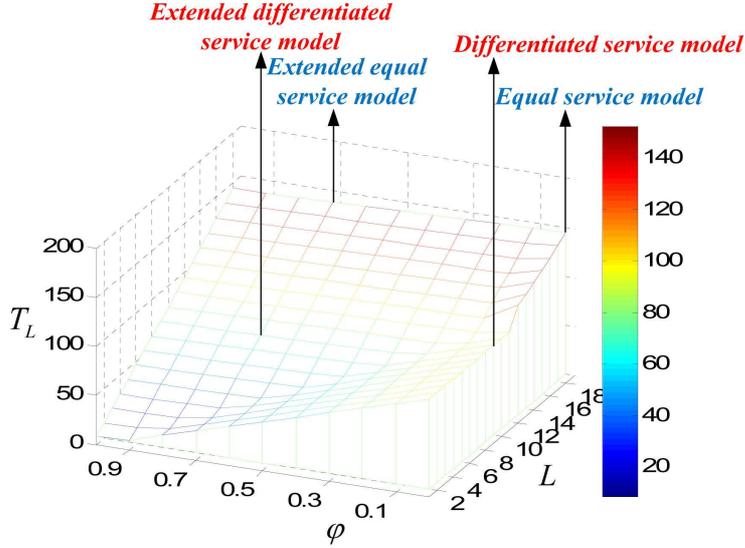}
\caption{\emph{Equal service} and \emph{differentiated service} models. ($\varphi \in [0,1]$)}
\label{numerical}
\end{figure}

\section{Conclusions}
In this paper, we modeled both the equal service process and the differentiated service process when peers' initial data distribution satisfies some special conditions (the \emph{UCP condition}), and derived exact expression of the minimum last finish time. Both the \emph{UCP condition} and the proposed models based on it have great theoretical and practical value.

\begin{itemize}
  \item The intrinsic relations among the initial data amount, the size of peer set and the minimum last finish time are revealed. Simply speaking, if the total data amount in all peers is small, the minimum last finish time is constrained by the source and cannot be further shortened by reducing the size of the first peer set. On the contrary, if the total data amount is large, the minimum last finish time can be much smaller than in the former case, and fewer peers can reach it.
  \item After the first peer set finishes its download, the data distribution of the second peer set still satisfies the \emph{UCP condition}. Therefore, we can further use our proposed extended models to provide equal service or differentiated service to the second peer set. Furthermore, we can provide arbitrary degree of differentiated service to a certain number of peers. Suppose there are $k$ sets of peers, then we can provide differentiated service from the first peer set to the $k$-th peer set in turn, the whole download process can be decomposed into $k-1$ nested differentiated service processes and one equal service process, whichever process can be characterized by our extended differentiated service model or equal service model.
\end{itemize}

Due to there are no limits for peers' upload bandwidth and the size of each peer set, our proposed models have wide applicability and high extensibility for peer-to-peer file sharing.

\section*{Acknowledgement}
This work is supported by the Next Generation Internet special of NDRC (Grant No. CNGI2008--112) and the National Key Technology R\&D Program (Grant No. 2009BAH43B02).
\section*{Appendix}
In this appendix we proof Theorem 1.

Let ${{U}_{r,L,N-L}}'(L,{{t}_{0}})$ denote $(1-\frac{1}{L}){{U}_{r,N-L}}(L,{{t}_{0}})+{{U}_{r,L}}(L,{{t}_{0}})$ in inequality (33).
Define a function $ur(l)$ such that:

\begin{equation}\label{eq:35}
ur(l)={{{U}_{r,l,N-l}}'(l,{{t}_{0}})}/{(l-1)}\;
\end{equation}

By simple calculation we have $ur(l)=\frac{U}{l-1}-\frac{{{F}_{a}}}{{{t}_{0}}}-\frac{{{C}_{N-l}}}{l(l-1)}$, then:

\begin{equation}\label{eq:36}
ur(l+1)-ur(l)=\frac{{{C}_{N-l}}-U}{l(l-1)}-\frac{{{C}_{N-(l+1)}}}{l(l+1)}<0
\end{equation}

If inequality (34) is true when $L=l+1$, we have $ur(l+1)\ge {{C}_{0}}$, therefore $ur(l)>{{C}_{0}}$ holds, i.e. inequality (34) is true when $L=l$. Obviously, it is also true when $L<l$.

${{U}_{r,L,N-L}}(L,{{t}_{0}})$ can be rewritten as ${{U}_{r,L,N-L}}(L,{{t}_{0}})=U-\frac{(L-1){{F}_{a}}+{{F}_{N-L}}}{{{t}_{0}}}$ by definition. Since ${{t}_{0}}\le (L'+1){{t}_{a}}$, we have:

\begin{equation}\label{eq:37}
{{U}_{r,L'+1,N-(L'+1)}}(L'+1,{{t}_{0}})\le {{U}_{r,L'+1,N-(L'+1)}}'(L'+1,{{t}_{0}})
\end{equation}

Combining (35) - (37), we have ${{{U}_{r,L'+1,N-(L'+1)}}(L'+1,{{t}_{0}})}/{L'}\;<{{{U}_{r,L',N-L'}}'(L',{{t}_{0}})}/{(L'-1}\;)$.
If inequality (32) is true when $L=L'+1$, we have ${{{U}_{r,L'+1,N-(L'+1)}}(L'+1,{{t}_{0}})}/{L'}\;\ge {{C}_{0}}$.therefore ${{{U}_{r,L',N-L'}}'(L',{{t}_{0}})}/{(L'-1}\;)>{{C}_{0}}$ holds, i.e. inequality (34) is true when $L=L'$. Obviously, it is also true when $L<L'$.

\end{document}